\documentclass[twocolumn,aps,prl,amsfonts]{revtex4}
\usepackage{latexsym}
\usepackage{amsmath}
\usepackage{amssymb}
\usepackage{graphicx,color,psfrag}

\newcommand{\afour}{|a_4 \rangle}
\newcommand{\aeight}{|a_8 \rangle}
\newcommand{\aeightfour}{|a_8^{(4)} \rangle}

\newcommand{\phasegate}{\Lambda(e^{i \pi/4})}
\newcommand{\notg}{ \begin{small}NOT\end{small}}
\newcommand{\cnot}{\begin{small}CNOT\end{small}}
\newcommand{\ccnot}{\begin{small}CCNOT\end{small}}
\newcommand{\order}{\mathcal{O}}
\newcommand{\cN}{\mathcal{N}}
\newcommand{\cL}{\mathcal{L}}

\begin{document}

\title{Resources Required for Topological Quantum Factoring}

\author{M. Baraban$^1$, N. E. Bonesteel$^2$, S. H. Simon$^3$}
\affiliation{$^1$ Department of Physics, Yale University, 217 Prospect Street, New Haven, Connecticut 06511 \\
$^2$ Department of Physics and National High Magnetic Field
Laboratory, Florida State University, Tallahassee, FL 32310 \\
$^3$ Rudolf Peierls Centre for Theoretical Physics, Oxford University, 1 Keble Road, Oxford OX1 3NP, UK}

\begin{abstract}
We consider a hypothetical topological quantum computer where the qubits are comprised of either Ising or Fibonacci anyons.  For each case, we calculate the time and number of qubits (space) necessary to execute the most computationally expensive step of Shor's algorithm, modular exponentiation.  For Ising anyons, we apply Bravyi's distillation method [S. Bravyi, Phys. Rev. A \textbf{73}, 042313 (2006)] which combines topological and non-topological operations to allow for universal quantum computation.  With reasonable restrictions on the physical parameters we find that factoring a 128 bit number requires approximately $10^3$ Fibonacci anyons versus at least $3\times10^{9}$ Ising anyons.  Other distillation algorithms could reduce the resources for Ising anyons substantially. 
\end{abstract}

\maketitle

Shor's algorithm is at the center of much of the excitement surrounding quantum computation.  Classically, the time to factor a number of length $\cL$ grows exponentially in $\cL$, but given a sufficiently large quantum computer, Shor's algorithm could be used to factor in polynomial time \cite{shor}.  Specifically, the most computationally expensive step of Shor's algorithm is modular exponentiation which scales as $\cL^3$.   Since internet security is based on the near impossibility of factoring large numbers, the ability to factor in polynomial time, or in other words, the existence of a sufficiently large quantum computer, would be of monumental importance.  In this letter, we will address the question of what \emph{sufficiently large} means for a topological quantum computer.

Many different systems have been proposed as the building blocks for a quantum computer known as quantum bits or qubits, but we will focus on topologically protected qubits which are created using non-Abelian particles \cite{nayak_rmp}.  Topological systems are particularly attractive candidates for quantum computation because of their natural resistance to decoherence.  Non-Abelian particles have the property that topological operations, or braiding the particles around each other at large distances, can rotate the system between its degenerate ground states.  The ground state degeneracy grows exponentially with the number of particles allowing groups of particles to store quantum information in the form of qubits \cite{kitaev_tqc,nayak_rmp}.  

The most commonly considered non-Abelian particle is the Majorana fermion, or Ising anyon, where it is most convenient to use four Ising anyons to form a single qubit.  Ising anyons are expected to be the excitations of the $\nu=5/2$ fractional quantum Hall state \cite{mr}.  Additionally, there have been proposals to create Ising anyons in Sr$_2$RuO$_4$ thin films \cite{dassarma_films}, cold atoms \cite{coldatoms}, and most recently in several varieties of strongly coupled spin orbit systems involving superconducting junctions \cite{junctions}.  While Ising anyons are the simplest example of a non-Abelian particle, braiding Ising anyons is not sufficient for universal quantum computation (UQC).  Bravyi has suggested a method for combining topological and non-topological operations to allow for UQC with Ising anyons \cite{bravyi}.  We will explore this method in detail below, but the basic strategy is to create entangled states using non-topological operations and then braid these states with the target qubits to perform gates that are not allowed topologically.  We find that when the time to prepare these entangled states is large compared to the time to run the algorithm, the number of qubits required to perform the algorithm scales approximately as the number of gates, $\cN$, which is proportional to $\cL^3$.    

A second type of non-Abelian particle are Fibonacci anyons which are expected to be the excitations of the $\nu=12/5$ fractional quantum Hall state \cite{read_rezayi} and exist in certain toy lattice models \cite{fiblattice}.  Here it is convenient to use three Fibonacci anyons to form a single qubit.  Since braiding Fibonacci anyons is sufficient for UQC, the number of Fibonacci anyons needed to factor a number of length $\cL$ scales as $\cL$ rather than $\cL^3$.  Practically, this means that factoring a 128 bit number requires approximately $10^3$ Fibonacci anyons rather than $3 \times 10^{9}$ Ising anyons.  While $10^{9}$ is a huge number, Ising anyons remain attractive as a possible platform for quantum computation because, as shown by Bravyi \cite{bravyi}, and seen explicitely below, there is a high error tolerance for the non-topological operations necessary to prepare the states.

To estimate the number of particles necessary for modular exponentiation, we will assume that all braid operations can be performed perfectly and that error only results from the error intrinsic in the gates themselves.  For Ising anyons, the error stems from the non-topological operations needed to prepare the entangled states, while for Fibonacci anyons, the length of the braid determines the accuracy of the gate.  In this paper, all errors will be stated as error probabilities (the \emph{square} of the amplitude), and for both Ising and Fibonacci anyons, the error per gate must be less than or on the order of $1/\cN$ where $\cN$ is the total number of gates \cite{error}.  Additionally, we assume that the state of the qubit can be measured efficiently and with negligible error (which can always be achieved by repetitive measurements).  The number of \notg, \cnot, and \ccnot\ gates required for efficient modular exponentiation is proportional to $\cL^3$ and was calculated precisely in Ref.~\cite{beckman_preskill}.  For $\cL=128$, the total number of gates is $\cN \approx 10^9$. 

\textit{\textbf{Ising Anyons:}}  
Bravyi's method to achieve UQC with Ising anyons uses non-topological operations to poorly approximate the one and two qubit states $\afour =  \frac{1}{\sqrt{2}} \left( |0\rangle + e^{i \pi/4} |1\rangle \right)$ and $\aeight =  \frac{1}{\sqrt{2}}\left( |0,0\rangle + |1,1\rangle \right)$.  These states are then distilled in a process that takes many states with large error to a single state with smaller error using only topological operations and measurements \cite{bravyi, bravyi_kitaev}.  States with arbitrarily small error can be produced by repeating the distillation process.  The purified $\afour$ and $\aeight$ states are then braided with the target qubits to implement the controlled $\pi/4$ phase gate, $\phasegate$, and \cnot\ gates.  The combination of \cnot\ and $\phasegate$ allows for UQC with Ising anyons \cite{chuang}.  

We will calculate how many qubits and how many operations are necessary to first distill the states and then execute the modular exponentiation algorithm.  Ref.~\cite{bravyi} carried out a similar calculation and found that to distill $\cN$ states, the number of operations and qubits necessary scaled as $\cN (\ln \cN)^3$.  This calculation assumed that the error in the initial states was asymptotically small, and in this limit, the distillation procedure is successful nearly $100 \%$ of the time.  Since the probability of a successful distillation vanishes as the initial error approaches an upper bound, the number of qubits required to distill one state depends strongly on the initial error.  Additionally, the calculation did not account for the possibility of reusing qubits or performing distillation rounds in parallel.  
In theory, one could imagine starting with enough qubits to distill all $\cN$ states simultaneously without qubit reuse and then performing the modular exponentiation.  This would minimize the time; however, as we will see, it would also require a gigantic number of qubits. 
In our calculation, we will not assume asymptotically small initial error and we will explore the balance between the time and space requirements by combining parallel operations with reusing qubits.  Initially, since $\cN$ distilled states are required to perform the algorithm, let us assume we have at least $\cN$ qubits to work with (which is already a large number) and we will attempt to perform the full distillation and algorithm with no more than this number (we will further examine this requirement below).

\begin{figure}[tb]
\includegraphics[width=0.4\textwidth]{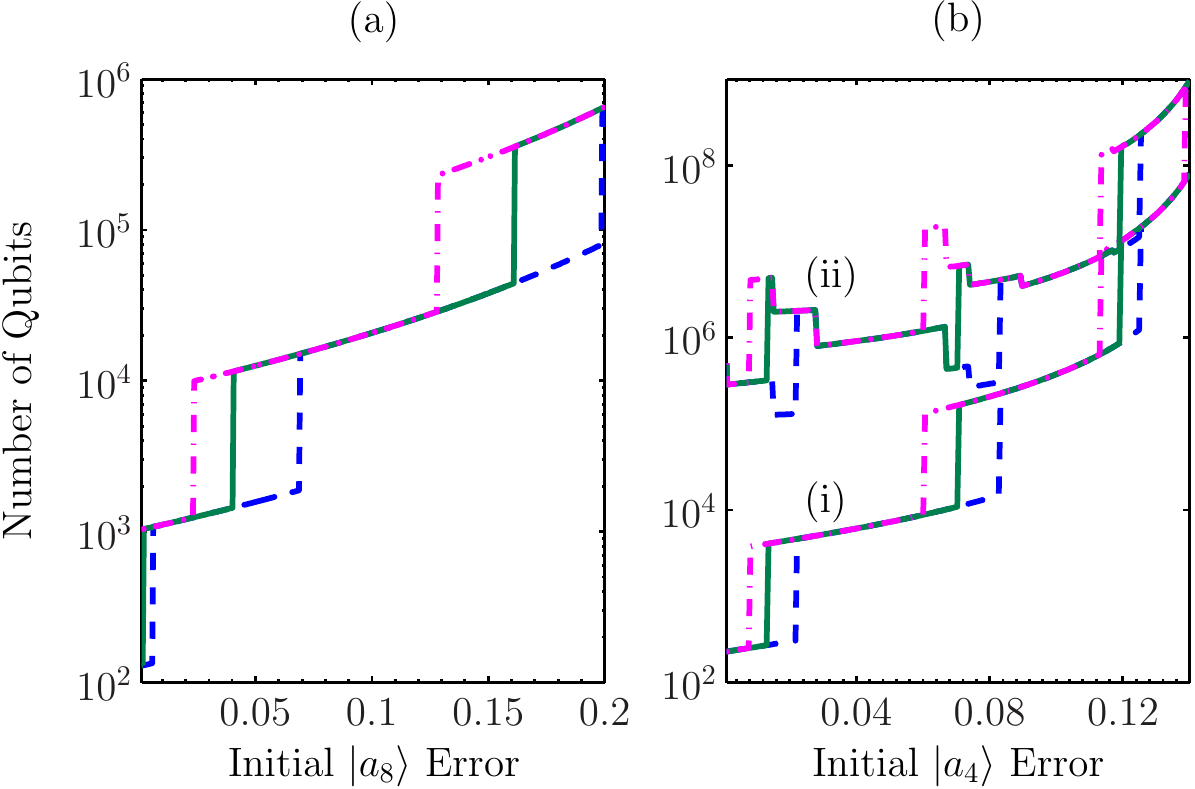}
\caption{(color online) Plot (a) shows the number of qubits required to create a single $\aeight$ state with final error of $10^{-9}$ (dashed), $10^{-11}$ (solid), and $10^{-13}$ (dash-dot).  The jumps indicate where the number of distillation levels increases by one, with plateau at 0.1 being four purification rounds.  Plot (b) shows the number of qubits to create a single $\afour$ state where (i)/(ii) is the qubits to make $\afour$/$\aeight$ states.  The initial error of the $\aeight$ states is taken to be 0.01, and the final errors for $\afour$ are the same as plot (a).}
\label{fig:qubits}
\end{figure}

The number of qubits necessary to distill a single $\aeight$ state is shown in Fig.~\ref{fig:qubits}(a) where $\aeight$ distillation is only successful when the initial error is less than $0.38$.  Notice that even for a relatively large initial error, the number of qubits to distill one $\aeight$ state is small compared to $\cN$ for $\cL =128$.  Given our attempt to limit the number of qubits, we choose to create $\cN$ $\aeight$ states using $\order(\cN)$ qubits.  Specifically, we will start with about $\cN$ poorly approximated $\aeight$ states and perform the distillation algorithm in parallel on all these initial states.  This will result in a small fraction of the initial qubits being converted into fully distilled $\aeight$ states.  The remainder of the qubits can be reinitialized to poor approximations of $\aeight$ and again distilled in parallel to purified $\aeight$ states.  By repeating this process, nearly all the initial qubits can be converted into fully distilled $\aeight$ states.  Note that the distillation process ends by measuring the qubits which are not part of the distilled state.  Since these qubits are no longer entangled with the purified state, they can easily be reused in a subsequent distillation.

$\afour$ distillation has an added complication because $\aeight$ states are required in the distillation process.  At each level of distillation, the $\aeight$ states must have an error at least as small as the final $\afour$ states.  The number of qubits to distill a single $\afour$ state is shown in Fig.~\ref{fig:qubits}(b), where the qubits needed for the $\afour$ and $\aeight$ states are plotted separately, and the maximum error for the initial $\afour$ state is $0.14$.  Notice the number of qubits needed for $\aeight$ states sometimes decreases as the initial error increases.  These decreases result from a technicality where less exact $\aeight$ states are required within the $\afour$ distillation round.  
To avoid confusion, we will not plot these nonmonotonicities in the future as the distillation can always be run assuming the larger error.

To distill many $\afour$ states, we will again choose to minimize the space requirements and use only $\order(\cN)$ qubits to distill $\cN$ $\afour$ states.   Since the number of qubits needed to make $\aeight$ states for a single $\afour$ distillation (we will call these states $\aeightfour$) approaches $\cN$ for $\cL = 128$, our distillation scheme will be to dedicate approximately $\cN$ qubits to making $\aeightfour$ states.
We will then make as many $\aeightfour$ states as possible, do the $\afour$ distillation, go back and reuse the qubits to make more $\aeightfour$ states, do another round of $\afour$ distillation and repeat this process until the $\afour$ states are fully distilled.  
$\afour$ distillation is slow compared to $\aeight$ distillation, so when considering the total resources required, there will be a one to one trade-off between space and time that depends on the number of qubits dedicated to $\aeightfour$ production.  

\begin{figure}[tb]
\includegraphics[bb=150 405 482 663,width=0.45\textwidth]{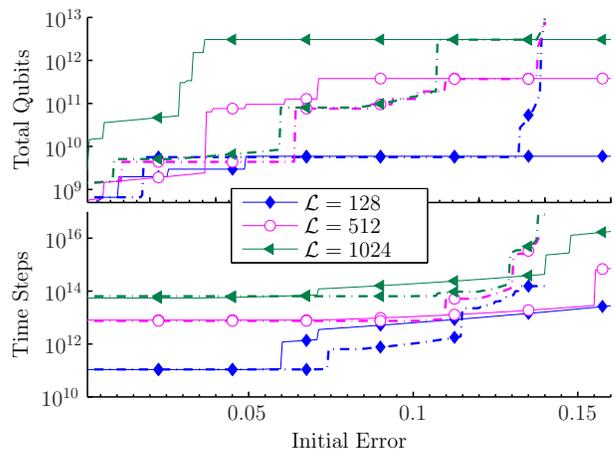} 
\caption{(color online) Total qubits and time steps necessary for modular exponentiation.  The connected (dashed) lines are for changing the initial error of the $\aeight$ ($\afour$) states while the $\afour$ ($\aeight$) initial error is held constant at 0.01.}
\label{fig:modexp}
\end{figure}

The results for how many qubits and how much time is required to perform modular exponentiation with Ising anyons are shown in Fig.~\ref{fig:modexp}.  We define the time to perform one braid as a time step, and our calculations assume that a measurement can also be performed in a single time step.  This assumption is probably unrealistic, but currently no model exists for estimating the time of a measurement.  Measurements account for between $5$ and $8\%$ of the operations while running the distillation and executing the \cnot\ and \ccnot\ gates with the remainder being topological braids.  If the time to perform a measurement is an order of magnitude longer than to perform a braid, then the total time will nearly double from that plotted in Fig.~\ref{fig:modexp} (this may be the case if repetitive measurement is required to obtain reliability).

\begin{figure}[tb]
\includegraphics[width=0.4\textwidth]{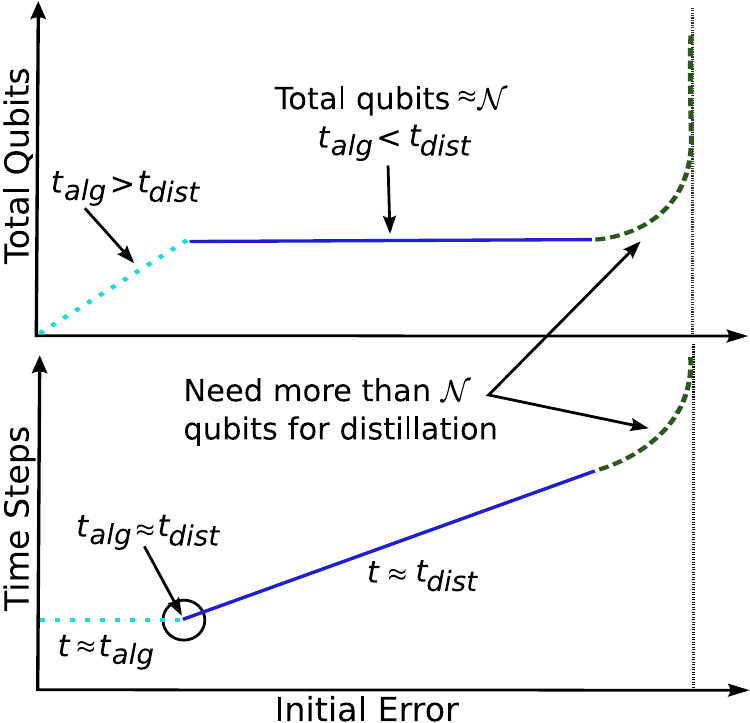}\\
\caption{(color online) Cartoon sketch to help clarify Fig.~\ref{fig:modexp}.  The three regimes which describe our results are when the time to run the algorithm, $t_{alg}$, is (a) greater than $t_{dist}$, (b) less than $t_{dist}$, and (c) when the number of qubits to distill a single state is comparable or large compared to $\cN$.  
These three regimes are seen in Fig.~\ref{fig:modexp} for $\cL =512$ while changing the initial error of $\aeight$ (solid line) for the approximate values (a) 0-0.07, (b) 0.07-0.35, and (c) 0.35-0.38 (not shown).}
\label{fig:cartoon}
\end{figure}

Fig.~\ref{fig:modexp} can be considered in three regimes which are sketched in Fig.~\ref{fig:cartoon}. When the initial error is small enough, $t_{dist}$, the time to distill all the required $\aeight$ and $\afour$ states as outlined above, is short compared to the time to run the algorithm, $t_{alg}$.  Rather than distilling the states all at once, we will minimize the number of qubits subject to the constraint that the total distillation time is
comparable to $t_{alg}$.  In this model, states are distilled, used to perform the algorithm, and then the qubits are reused to distill more states and continue the algorithm, and so on until the algorithm is completed.  For any given $\cL$, there always exists an initial error small enough such that the time to run the algorithm dominates, so state distillation does not need to change the scaling of the total time.  In Figs.~\ref{fig:modexp} and \ref{fig:cartoon}, this regime is seen where the time is nearly flat and the number of qubits is increasing.

For larger values of the initial error, $t_{dist}$ becomes large compared to $t_{alg}$, and we choose to use $\order(\cN)$ qubits to distill all the states at once as described previously.  In this case, the time does not depend directly on $\cL$, but rather on the number of distillation rounds necessary to fully purify the $\afour$ and $\aeight$ states.   Hence, the time is nearly independent of $\cL$ across a wide range of values and increases as the initial error increases.  We choose the total number of qubits to scale as $\cN$, but there is approximately a one to one trade off between space and time.  Conceivably, the number of qubits could be significantly reduced, but the total time would increase comparably.  Additionally, for any initial error, there is always an $\cL$ large enough to return us to the previous region where $t_{dist} < t_{alg}$. 

Finally, as the initial error approaches its upper bound, the number of qubits to distill a single state becomes comparable and eventually exceeds $\cN$.  Once this happens, both the total number of qubits and total distillation time diverge.  Note that the time does not exhibit a strong divergence in Fig.~\ref{fig:modexp} as the initial error of the $\afour$ states is increased because the initial $\aeight$ error remains small.  The details of Fig.~\ref{fig:modexp} depend on the exact distillation scheme, but the qualitative results sketched in Fig.~\ref{fig:cartoon} are more universal.

Using the results from Ref.~\cite{baraban} and the distillation scheme in Fig.~\ref{fig:modexp}, we can estimate the size of a $\nu = 5/2$ sample and the total time necessary to perform modular exponentiation when $\cL=128$.  For approximately $10^9$ gates, each gate must have error $\lesssim 10^{-9}$.  Taking the initial $\afour$ and $\aeight$ error to be $ 0.01$, we need approximately $10^{11}$ time steps and $3\times10^{9}$ quasiparticles (qp's) to create all the states and perform the algorithm (see Fig.~\ref{fig:modexp}).
To manipulate the qp's, we apply an electric field with magnitude much less than $\sim \Delta/(e^* \ell^*) $ to avoid particle-hole pair creation where $\Delta$ is the gap of the 5/2 state, $e^*=e/4$ is the qp charge  and  $\ell^* = 2 \ell$ is the effective magnetic length.  This
results in a maximum $E \times B$ drift velocity of  order $\Delta \ell / \hbar$.   Using
the decay length from Ref.~\cite{baraban}, and assuming $\Delta = 1$K,
we find that the qp's need to be separated by at least $100 \ell$ and that the maximum step rate is 30 MHz. Modern single
electron pumps can function at a rate of nearly 20 MHz
with error rates as low as 15 per $10^9$ \cite{keller}, so achieving
30 MHz with comparably low error seems plausible. At
this step rate, the calculation would take approximately
$3 \times 10^3$s on a sample that is at least 10cm x 10cm.  While we can trade some
amount of space for time, if one were to reduce the space by more than
a few orders of magnitude, the runtime would become sufficiently long
that classical computers could potentially compete.
Parameters will differ and may be more favorable for other potential systems of Ising anyons \cite{dassarma_films,coldatoms,junctions}.

\textit{\textbf{Fibonacci Anyons:}}
Computation with Fibonacci anyons is in many ways much simpler than using Ising anyons.  Since braiding Fibonacci anyons is sufficient for UQC, we only need to find a braid to implement the desired gate.  Additional entangled states to act on the target qubits are not necessary, so the space needed to perform modular exponentiation will be $\order(\cL)$, the length of the number to factor.  Further, only $\order(\cL)$ measurements are required at the end of the calculation.  However, Fibonacci anyons do not naturally implement \notg, \cnot, or \ccnot\ gates \cite{freedman}, so the challenge is to find a braid which approximates the desired gate to the necessary level of accuracy.

Brute force searches for braids have found that gate error becomes exponentially small as the braid length is increased linearly \cite{hormozi}; however due to computational difficulty, the longest 
brute force braid available
is about 80 steps with an error of about $10^{-10}$.  The accuracy of any braid can be improved using the Solovay-Kitaev (SK) algorithm \cite{chuang}.  With each iteration of SK, the error improves as $\epsilon_1 \sim c \epsilon_0^{3/2}$ and the braid length increases by a factor of 5.  Therefore, we can again construct gates with arbitrarily small error, but at the expense of the braid length growing as $5^n$, where $n$ is the number of SK iterations.  (Other schemes to obtain longer and more accurate braids may replace or be combined with SK beyond where brute force searches are feasible \cite{quantumhashing}.)

\begin{figure}[tb]
\includegraphics[width=0.4\textwidth]{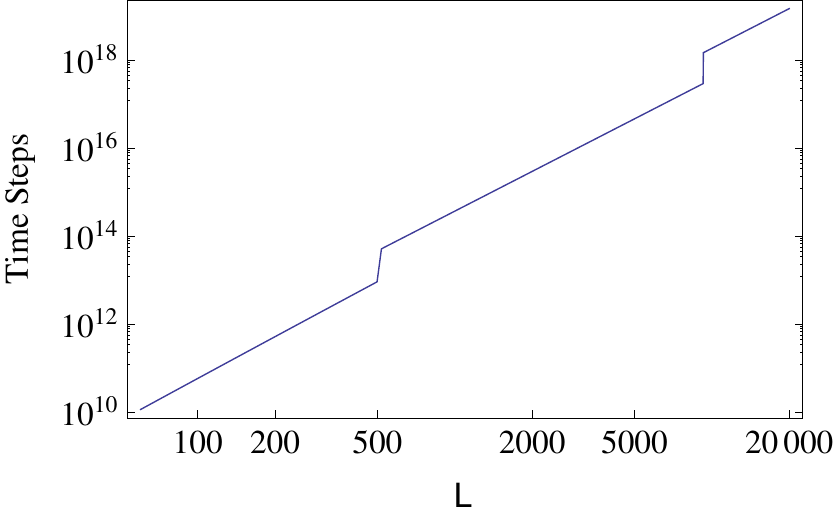}
\caption{Number of time steps needed to complete modular exponentiation with Fibonacci anyons.  Each jump is where an additional SK iteration is required, and for $\cL \lesssim 500$, the braid is determined solely by the brute force search method.}
\label{fig:fib}
\end{figure}

Fig.~\ref{fig:fib} shows the time to complete modular exponentiation using Fibonacci anyons as a function of $\cL$. 
The total space scales as $\cL$ and is $2 \cL + 3$ for this specific implementation of modular exponentiation \cite{beckman_preskill}.  For $\cL = 128$, modular exponentiation requires 259 qubits (777 Fibonacci anyons).  The time will be proportional to the braid length per gate times the number of gates, which is approximately $10^{11}$ time steps.  Assuming a comprable minimum distance between qp's in the $\nu = 12/5$ state as the 5/2 state, the gap in the $12/5$ state restricts the maximum step rate to about 3 MHz, so this computation would take on the order of $3 \times 10^4$ seconds.

To summarize, we explore the space and time requirements of Bravyi's distillation technique for Ising anyons. We find a
good balance by producing $\cN$ nontopological gates using $\order(\cN)$ qubits. For this
to succeed, the initial error in the $\afour$ and $\aeight$ states must be small enough such that the number of qubits needed to distill a single $\aeight$ or $\afour$ state is small compared to $\cN$.  When the time to run the algorithm is small compared to the time to distill states, we can reduce the space even further by distilling the states in batches while running the algorithm.  We note that we have made certain assumptions concerning the trade-offs between space and time which we believe are appropriate and would give the best possible outcome in a realistic system.  However, other choices can be made, and the results can be worked out from the details we provide.

Note Added:  Since the completion of this work we have learned of an unpublished method that allows for \cnot\ without $\aeight$ distillation \cite{bonderson}.  Analysis of this new algorithm is beyond the scope of the current work, but rough estimates suggest that this new scheme could reduce $t_{dist}$ by a factor of $\sim 10^5$ compared to the example presented here.  

The authors are grateful to M. Freedman and P. Bonderson for informing us of Ref.~\cite{bonderson}.  M.B. is supported by NSF Grant No. DMR-0603369, and N.E.B is supported by US DOE Grant No. DE-FG02-97ER45639.

\vspace*{-10pt}

\vspace*{-10pt}


\begin{thebibliography}{17}
\providecommand{\natexlab}[1]{#1}
\providecommand{\url}[1]{\texttt{#1}}
\providecommand{\urlprefix}{URL }
\providecommand{\bibAnnoteFile}[1]{%
  \IfFileExists{#1}{\begin{quotation}\noindent\textsc{Key:} #1\\
  \textsc{Annotation:}\ \input{#1}\end{quotation}}{}}
\providecommand{\bibAnnote}[2]{%
  \begin{quotation}\noindent\textsc{Key:} #1\\
  \textsc{Annotation:}\ #2\end{quotation}}
\providecommand{\eprint}[2][]{\url{#2}}

\bibitem{shor} P.~Shor, \textit{Proceedings of the 35th Annual Symposium on Foundations of Computer Science}, edited by S. Goldwasser 
(IEEE Computer Society, Los Alamitos, CA, USA, 1994), pp. 124Ð134. 

\bibitem{nayak_rmp} C.~Nayak et al., Rev.~Mod.~Phys. \textbf{80}, 1083 (2008). 

\bibitem{kitaev_tqc} A.~Kitaev, Ann.~Phys. \textbf{303}, 2 (2003). 

\bibitem {mr} G.~Moore and N.~Read, Nucl.~Phys.~B \textbf{360}, 362 (1991). 

\bibitem {dassarma_films} S.~Das~Sarma, C.~Nayak, and S.~Tewari, Phys.~Rev.~B \textbf{73}, 220502 (2006). 

\bibitem {coldatoms} V.~Gurarie, L.~Radzihovsky, and A.~V.~Andreev, Phys. Rev. Lett. \textbf{94}, 230403 (2005); S.~Tewari et al., Phys. Rev. Lett. \textbf{98}, 010506 (2007);  N.~R.~Cooper and G.~V.~Shlyapnikov, Phys.~Rev.~Lett. \textbf{103}, 155302 (2009).

\bibitem {junctions} L.~Fu and C.~Kane, Phys.~Rev.~Lett. \textbf{100}, 96407 (2008); J.~Sau et al., arXiv:0907.2239 (2009); P.~A.~Lee, arXiv:0907.2681 (2009). 

\bibitem {bravyi} S.~Bravyi, Phys.~Rev.~A \textbf{73}, 042313 (2006). 

\bibitem {read_rezayi} N.~Read and E.~Rezayi, Phys.~Rev.~B \textbf{59}, 8084 (1999); E.~Rezayi and N.~Read, Phys.~Rev.~B \textbf{79}, 075306 (2009). 

\bibitem {fiblattice} M.~Levin and X.~G.~Wen, Phys.~Rev.~B \textbf{71}, 45110 (2005); P.~Fendley, Ann. Phys. \textbf{323}, 3113 (2008). 

\bibitem {error} This model assumes uncorrelated and phase coherent random errors. Additionally, for Fibonacci anyons the errors must add incoherently.  If any of these conditions are not met, a worst case situation would require gates with accuracy of $\order(1/\cN^2$). 

\bibitem {beckman_preskill} D.~Beckman et al., Phys.~Rev.~A \textbf{54}, 1034 (1996). 

\bibitem {chuang} M.~A.~Nielsen and I.~L.~Chuang, \textit{Quantum Computing and Quantum Information} (Cambridge University Press, Cambridge, 
England, 2000).

\bibitem {bravyi_kitaev} S.~Bravyi and A.~Kitaev, Phys.~Rev.~A \textbf{71}, 022316 (2005). 

\bibitem {baraban} M.~Baraban et al., Phys.~Rev.~Lett. \textbf{103}, 076801 (2009). 

\bibitem{keller} M.~W.~Keller et al., Science \textbf{285}, 1706 (1999).

\bibitem{freedman} M.~H.~Freedman and Z.~Wang, Phys.~Rev.~A \textbf{75}, 32322 (2007).

\bibitem {hormozi} L.~Hormozi et al., Phys.~Rev.~B \textbf{75}, 16 (2007). 

\bibitem {quantumhashing} M.~Burrelloar et al., arXiv:0903.1497 (2009); R. Mosseri, J. Phys. A: Math. Theor. \textbf{41} 175302 (2008).

\bibitem {bonderson} P.~Bonderson et al., unpublished.

\end{thebibliography}
\end{document}